\documentclass{article}
\usepackage{amsmath}
\usepackage{amssymb}
\usepackage{bbm}

\begin{document}

%\markboth{S. Kuwata}
%{Spin-dependent conformally invariant integral transform}


\title{Spin-dependent conformally invariant integral transform}

\author{Seiichi Kuwata \\
Graduate School of Information Sciences, Hiroshima City University\\
3-4-1 Ozuka-higashi, Asaminami-ku, Hiroshima 731-3194, Japan}

\maketitle

\begin{abstract}
As an extension of the Hilbert transform, which characterizes the spacetime conformally invariant integral transform,
we show that, based on the Bhabha theory,
a spin-dependent conformally invariant integral transform
can be written as a product of the spin-independent integral transform and the Casimir operator of the corresponding conformal group.
In contrast to an ordinary context, where spacetime translation symmetry is assumed,
we introduce an intrinsic momentum operator,
by which the Casimir operator turns out to be spacetime dependent
(spin-orbit coupling, Pauli-Lubanski pseudo vector, and the like).
We also show that the physical state annihilated by the intrinsic momentum operator represents a spin-$s$ state.
\end{abstract}

%\keywords{Bhabha theory; Casimir; conformal spinor; spin-orbit; twister.}

% Introduction
\section{Introduction}

Conformal symmetry~\cite{conformal} has many applications in general relativity,
string theory, statistical physics, and also in pure and applied mathematics.
It is known that the spacetime conformally invariant integral transform $\mathcal H_0: L^2 (\mathbb R^n) \rightarrow L^2 (\mathbb R^n)$
is given by such whose integral kernel 
is the Riesz potential~\cite{riesz}.
If we take account of a spin-dependent conformal transformation,
it is natural to ask what the spin-dependent conformally invariant integral transform $\mathcal H$ is like.
To answer this question,
we should clarify the spin-dependent conformal transformation.
In an ordinary context, Poincar\'e symmetry is respected~\cite{conformal,lagu,efim}.
In such a case, however, we find that the $\mathcal H_0$ cannot be a solution to the $\mathcal H$ as a particular solution.
Instead, we adopt an alternative approach, based on the Bhabha theory~\cite{bhabha}.
Introducing an intrinsic momentum operator $\pi_\mu$, which is analogous to the twister algebra~\cite{penrose},
we show that the $\mathcal H_0$ turns out to be a particular solution to $\mathcal H$.
Furthermore, we find that the general solution to $\mathcal H$ can be written as the product of $\mathcal H_0$ with the Casimir operator of the corresponding conformal group.
To recover the translational invariance of a physical state $| \psi_{\rm phys} \rangle$,
we introduce a projection ${\sf P}$ such that $| \psi_{\rm phys} \rangle = {\sf P} | \psi \rangle$ so as to satisfy $\pi_\mu | \psi_{\rm phys} \rangle = 0$.

% Integral transform
\section{Conformally invariant integral transform}

In this section, we deal with a conformally invariant integral transform.
The conformal algebra is given by
\begin{align}
[D, \, L_{\mu \nu}] & = 0, \qquad [ P_\mu, \, P_\nu] = 0 = [K_\mu, \, K_\nu] , \notag \\
[ D, \, P_\mu] &= {\rm i} P_\mu, \qquad [ D, \, K_\mu] = - {\rm i} K_\mu, \notag \\
[v_\rho, \, L_{\mu \nu} ] &= {\rm i} \left( g_{\rho \mu} v_\nu - g_{\rho \nu} v_\mu \right) \qquad ( v_\rho = P_\rho, K_\rho), \notag \\
[K_\mu, \, P_\nu] &= 2 {\rm i} \left( g_{\mu \nu} D - L_{\mu \nu} \right), \notag \\
[L_{\mu \nu}, \, L_{\rho \sigma} ] & = {\rm i} \left( g_{\mu \sigma} L_{\nu \rho} + g_{\nu \rho} L_{\mu \sigma} - g_{\mu \rho} L_{\nu \sigma} - g_{\nu \sigma} L_{\mu \rho} \right),
\label{eq:algebra}
\end{align}
where $D, P_\mu, K_\mu$, and $L_{\mu \nu}$ represent the
dilatation, momentum, special conformal (or reciprocal), and angular momentum operators, respectively.
Note that (\ref{eq:algebra}) is invariant under
\begin{align}
( D, \, P_\mu, \,  K_\mu, \, L_{\mu \nu} ) \rightarrow ( - D, \, K_\mu, \,  P_\mu, \, L_{\mu \nu} ),
\label{eq:invariant}
\end{align}
where we have used $L_{\nu \mu} = - L_{\mu \nu}$.

% Spin-indep. case
\subsection{Spin-independent case}

To begin with, we deal with a spin-independent conformally invariant integral transform.
For a $n$-dimensional spacetime, with the metric $g_{\mu \nu} = {\rm diag} \, (-1,1,1,\ldots, 1)$,
the conformal generators are given by
\begin{align}
D &= Q_\mu  P^\mu - \frac{\rm i}{2} d \qquad (d  \in  \mathbb C), \notag \\
K_\mu &= - (Q \cdot Q)  P_\mu + 2 Q_\mu D  \qquad \left ( Q \cdot Q = Q_\mu Q^\mu \right ), \notag \\
L_{\mu \nu} &= - Q_\mu P_\nu + Q_\nu P_\mu,
\label{eq:D=QP}
\end{align}
where
$Q_\mu$ and $P_\mu$, satisfying $[ Q_\mu, \, P_\nu ] = {\rm i} g_{\mu \nu} \mathbbm 1$, represent the position and momentum operators, respectively.
For the present purpose, we will find that we should choose the parameter $d$ ($\frac{d}{2}$ is known as a conformal degree) as $n$.
Note that the substitution (\ref{eq:invariant}) implies the transformation of $Q_\mu$ as
\begin{align*}
Q_\mu \rightarrow \widetilde{Q}_\mu := - \frac{Q_\mu}{Q \cdot Q},
\end{align*}
so that the translational property of the momentum operator as
${\rm e}^{{\rm i} a \cdot P} Q_\mu {\rm e}^{-{\rm i} a \cdot P} = Q_\mu + a_\mu$ ($a \cdot P = a^\mu P_\mu, \; a_\mu \in \mathbb R^n$) corresponds to
\begin{align*}
{\rm e}^{{\rm i} b \cdot K} \widetilde{Q}_\mu  {\rm e}^{-{\rm i} b \cdot K} = \widetilde{Q}_\mu + b_\mu \qquad ( b \cdot K = b_\mu K^\mu, \; b_\mu \in \mathbb R^n),
\end{align*}
which holds for all $d \in \mathbb C$.

Now we obtain an integral transform $\mathcal H: L^2 (\mathbb R^n) \rightarrow L^2 (\mathbb R^2)$ which is commutative with all the conformal generators.
Let $H (x,y): \mathbb R^n \times \mathbb R^n \rightarrow \mathbb C$ be the corresponding integral kernel.
Since we can derive the Lorentz invariance and the dilatation invariance from
$[ \mathcal H, \, P_\mu] = 0 = [ \mathcal H, \, K_\nu ]$,
together with $[ P_\mu, K_\nu] = 2 {\rm i} ( g_{\mu \nu} D - L_{\mu \nu})$ and
the Jacobi relation $[A, \, [B, \, C] ] + [B, \, [C, \, A] ] + [C, \, [A, \, B] ] = 0$,
we concentrate on the conditions $[ \mathcal H, \, P_\mu] = 0 = [ \mathcal H, \, K_\nu ]$.
Using the integration by parts (neglecting the boundary value at infinity),
we obtain
\begin{align*}
\left ( \forall \mu, \; [ \mathcal H, \, P_\mu ] = 0 \right)  & \Longrightarrow
\left( H (x,y) = \psi_{\rm p} (\ldots, z_\mu, \ldots), \quad z_\mu = x_\mu - y_\mu \right), \\
\left ( \forall \mu, \; [ \mathcal H, \, K_\mu ] = 0 \right)  & \Longrightarrow
\left( H (x,y) = W (x,y) \psi_{\rm k} ( \ldots, w_{\mu}, \ldots), \quad w_\mu = \widetilde{x}_\mu - \widetilde{y}_\mu \right), %\frac{x_\mu}{x \cdot x} - \frac{y_\mu}{ y \cdot y} \right),
\end{align*}
where
%$z_\mu = x_\mu - y_\mu$,
%$w_\mu = \frac{x_\mu}{x \cdot x} - \frac{y_\mu}{ y \cdot y}$,
$W (x,y) = (x \cdot x)^{- \frac{d}{2}} ( y \cdot y)^{\frac{d}{2} - n}$, and
$\psi_{\rm p}, \psi_{\rm k}: \mathbb R^n \rightarrow \mathbb C$ are certain functions to be determined.
Note that we have the following scaling property:
%
%\begin{align*}
$W ( \lambda x, \lambda y) = \lambda^{-2n} \, W (x,y) \; ( \lambda > 0)$.
%\end{align*}
%
Although we can derive the Lorentz invariance of $\mathcal H$ from
$[ \mathcal H, \, P_\mu] = 0 = [ \mathcal H, \, K_\nu ]$,
%together with $[ P_\mu, K_\nu] = 2 {\rm i} ( g_{\mu \nu} D - L_{\mu \nu})$ and
%the Jacobi relation $[A, \, [B, \, C] ] + [B, \, [C, \, A] ] + [C, \, [A, \, B] ] = 0$,
we use the Lorentz invariance of $\mathcal H$ explicitly.
Solving the condition $[ \mathcal H, \, L_{\mu \nu}] = 0$,
we find that $H (x,y)$ is given by a function of $\vec{x}_{\mu \nu} \cdot \vec{x}^{\mu \nu}$, $\vec{y}_{\mu \nu} \cdot \vec{y}^{\mu \nu}$, and
$\vec{x}_{\mu \nu} \cdot \vec{y}^{\mu \nu}$, where $\vec{x}_{\mu \nu} = (x_\mu, x_\nu)$, $\vec{y}_{\mu \nu} = (y_\mu, y_\nu)$, and the center dot $(\cdot)$ in
$\vec{a} \cdot \vec{b}$ represents a two-dimensional inner product as $a_1 b^1 + a_2 b^2$ for $\vec{a} = (a_1, a_2)$ and $\vec{b} = (b^1, b^2)$.
Thus the condition $[ \mathcal H, \, L_{\mu \nu}] = 0$ for all $\mu, \nu$ implies that $H (x,y)$ can be written as a function of
$x \cdot x$, $y \cdot y$, and $x \cdot y$,
which indicates that $\psi_{\rm p}$ and $\psi_{\rm k}$ depend on $z \cdot z$ and $w \cdot w$, respectively,
so that we can write $\psi_{\rm p}$ and $\psi_{\rm k}$ as
$\psi_{\rm p} = \phi_{\rm p} (z \cdot z)$ and $\psi_{\rm k} = \phi_{\rm k} ( w \cdot w)$, where $\phi_{\rm p}, \phi_{\rm k}: \mathbb R \rightarrow \mathbb C$.
Considering that we have $w \cdot w = \frac{z \cdot z}{( x \cdot x) ( y \cdot y)}$, we find that the condition $\phi_{\rm p} (z \cdot z) = W (x,y) \phi_{\rm k} ( w \cdot w)$ leads to
\begin{align*}
d = n, \qquad \phi_{\rm p} (x) = \phi_{\rm k} (x) = c \, x^{-\frac{n}{2}} \quad (c \in \mathbb C).
\end{align*}
In this case, the dilatation invariance of $\mathcal H$, that is, $H ( \lambda x, \lambda  y) = \lambda^{-n} H (x,y) \; (\lambda > 0)$ is satisfied,
as is expected from $[ K_\mu, \, P_\mu] = 2 {\rm i} g_{\mu \mu} D$.
Note that for $d = n$, $D$ and $K_\mu$ turn out to be (formally) hermitian in the sense that
%
%\begin{align*}
$\left . D \right |_{d=n} = \frac{1}{2} ( \hat{D} + \hat{D}^\dag )$,
$\left. K_\mu \right |_{d=n} = \frac{1}{2} ( \hat{K}_\mu + \hat{K}^\dag_\mu )$,
%\end{align*}
%
where $\hat{D} = \left . D \right |_{d=0}$ and $\hat{K}_\mu = \left . K_\mu \right |_{d=0}$.

So far, we have found that the particular limit of Riesz potential gives the integral kernel for the conformally invariant integral transform as
$H (x,y) = c \left[ ( x-y) \cdot (x-y) \right]^{- \frac{n}{2}}$ (the Riesz potential is not in general dilatation invariant).
Here we mention some remarks.
$H(x,y)$ has a singularity at $( x - y ) \cdot (x-y) = 0$.
To avoid the singularity, one stratedy is to take the Cauchy principal value, as in the case of $n=1$ (the ordinary Hilbert transform).
Another method is to adopt the Feymann's ${\rm i} \epsilon$ prescription,
% $\frac{1}{ (x - y) \cdot (x-y) - {\rm i} \epsilon}$,
as in the case of $n=2$, where
the Fourier transform of the Green function for d'Alembertian is modified as$\frac{1}{ - x_0^2 + x_1^2} \rightarrow \frac{1}{ - x_0^2 + x_1^2 - {\rm i} \epsilon}$.

Even if the singularty at $( x - y ) \cdot (x-y) = 0$  is eliminated,
$H (x,y)$ is not integrable with respect to $x$ (or $y$), that is, $\int_{\mathbb R^n} | H (x,y) | \, {\rm d} x $ cannot be finite.
Thus, the Fourier transform of $H (x,y)$ concerning $x$ (or $y$) proves to be conditionally convergent.
For example, we consider the Fourier transform of $\frac{1}{-x^2 + y^2 - {\rm i} \epsilon}$ ($x, y \in \mathbb R$) in several ways of integration order.
Let $f_0, f_1, f_{\sigma}, f_{\tau}: \mathbb R^2 \rightarrow \mathbb C$ be given by
\begin{align}
f_{0} & = \int_{\mathbb R} 
\left[  \int_{\mathbb R} F \, {\rm d} \xi  \right ] \!
{\rm d} \eta, \qquad
f_{1}  =  \int_{\mathbb R}
\left[  \int_{\mathbb R} F \, {\rm d} \eta  \right ] \!
{\rm d} \xi, \notag \\
f_{\sigma} & = \int_{\mathbb R} 
\left[  \int_{\mathbb R} \widetilde{F}  \, {\rm d} s  \right ] \!
{\rm d} t, \qquad
f_{\tau}  = \int_{\mathbb R} 
\left[  \int_{\mathbb R} \widetilde{F}  \, {\rm d} t  \right ] \!
{\rm d} s,
\label{eq:f0=int}
\end{align}
where
$F = F(x, y; \xi, \eta) = \frac{ {\rm e}^{- {\rm i} x \xi } {\rm e}^{ {\rm i} y \eta} }{ - \xi^{2} + \eta^{2} - {\rm i} \epsilon }$, and
$\widetilde{F}$ is given by rewriting $F$ using $s = \xi + \eta$ and $t = \xi - \eta$ as
$\widetilde{F} (x,y; s,t) {\rm d} s \wedge {\rm d} t = F(x, y; \xi, \eta) {\rm d} \xi \wedge {\rm d} \eta$,
% ($s = \xi + \eta, t = \xi - \eta$),
so that we have
%
%\begin{align*}
$\widetilde{F} = \widetilde{F} (x,y; s, t)  = \frac{1}{2} \frac{ {\rm e}^{- \frac{1}{2} {\rm i} (x-y) s } {\rm e}^{ - \frac{1}{2} {\rm i} (x+y) t} }{ - st - {\rm i} \epsilon }$.
%\end{align*}
%
By definition, it is straightforward that $f_i = f_i (x,y) \; (i  \in \{ 0,1, {\sigma}, {\tau} \})$
satisfies the following relations:
\begin{align}
f_1 (x,y) = - \overline{f_0 (y,x)}, \qquad f_{\tau} (x,y) = f_{\sigma} (x,-y),
\label{eq:f1=-f0}
\end{align}
where $\overline{z}$ represents the complex conjugate of $z \in \mathbb C$.
Let $X_i$ and $Y_i$ be the real and imaginary parts of $f_i$, respectively.
Then, the explicit calculation yields
\begin{align}
X_{\sigma} = X_{\tau} = \frac{1}{2} \left( X_0 + X_1 \right ), \qquad
Y_0 = Y_1 = Y_{\sigma} + Y_{\tau},
%f_0 = X_0 + {\rm i} Y, \qquad f_1 = X_1 + 
%
%f_{\sigma} = f_{\tau} = \frac{1}{2} ( f_0 + f_1), \qquad f_0 = X + {\rm i} Y,
\label{eq:Xs=Xt}
\end{align}
where $X_j = X_j (x,y)$ ($j = 0,1$) and $Y_k = Y_k (x,y)$ $(k = \sigma, \tau)$ are given by%and $X_1 = X_1 (x,y)$ as
\begin{align*}
X_0 (x,y) & = \frac{\pi^2}{2} \left [ {\rm sgn} \, ( | x | + | y | ) + {\rm sgn} \, ( | x | - | y | )  \right ], \qquad
X_1 (x,y)  = - X_0 (y,x), \\
Y_{\sigma} (x,y) & = - \pi \left ( \gamma + \log | x + y | + \log \frac{\epsilon}{2} \right ), \qquad \qquad
Y_{\tau} (x,y) = Y_{\sigma} (x, - y),
\end{align*}
with ${\rm sgn} \, (x) = 1 \; (x>0)$, $0 \; (x=0)$, $-1 \; (x < 0)$,
and $\gamma \; (= 0.5772\ldots)$ representing the Euler's constant.
Thus, we find that
$X_{\sigma} (x,y) = X_{\tau} (x,y) = \frac{\pi^2}{2} {\rm sgn} \, ( | x | - | y | )$,
which can be rewritten as $\frac{\pi^2}{2} {\rm sgn} \, ( x^2 - y^2) = \frac{\pi^2}{2} {\rm sgn} \, ( x+ y ) \, {\rm sgn} \, (x-y)$.
In a similar way,
we obtain $Y_0 (x,y) = Y_1 (x,y) = - \pi ( 2 \gamma + \log | x^{2} - y^{2} | + 2 \log \frac{\epsilon}{2} )$.
%- \pi (\gamma + \log | x - y | + \log \epsilon$,
%where $\gamma$ represents the Euler's constant ($= 0.5772156649\ldots$).
%
In calculating $Y_{\sigma}$, we have regularized the integration over $t$ as
$\int_0^\infty \frac{1}{t} \cos \frac{1}{2} (x+y) \, {\rm d} t \rightarrow 
\int_{0}^\infty \frac{1}{t \pm {\rm i} \epsilon} \cos \frac{1}{2} (x+y) \, {\rm d} t$
(either sign in front of $\epsilon$ is possible, leading to the same result),
and have used the formula
$\int_0^\infty \frac{t \cos ax}{t^2 + b^2}  \, {\rm d}t = - \frac{1}{2} \left [ {\rm e}^{ - ab } \, {\rm Ei} \, (ab) + {\rm e}^{ab} \, {\rm Ei} \, (-ab) \right ]$ ($a, {\rm Re} \, (b) > 0$)~\cite{formula},
where ${\rm Ei} (x) = \int_{- \infty}^x \frac{{\rm e}^t}{t} \, {\rm d} t
%together with the asymptotic expansion
= \gamma + \log | x | + x +  O (x^2) \; (x \rightarrow 0)$.
%for ${\rm Ei} (x) = \int_{- \infty}^x \frac{{\rm e}^t}{t} \, {\rm d} t$.
Note that all the $f_i$'s ($i \in \{ 0,1,{\sigma}, {\tau} \}$) satisfy the same differential equation
$(\partial_x^{\, 2} - \partial_y^{\, 2}) f_i (x,y) = (2\pi)^2 \delta (x) \delta (y)$.
In this sense,
the conditional convergence of the Fourier transform of $H(x,y)$ may not cause a serious problem.

% Spin-dep. case
\subsection{Spin-dependent case}

In this subsection, we study the spin-dependent conformal invariant integral transform.
Let $L_{\mu \nu}$ be modified as
$L_{\mu \nu} \rightarrow L'_{\mu \nu} = L_{\mu \nu} + s_{\mu \nu}$,
where $s_{\mu \nu}$ represents the spin operator satisfying the same commutation relation as $L_{\mu \nu}$ in the last relation of (\ref{eq:algebra}).
Accordingly, we introduce $\Delta, \pi_\mu$, and $\kappa_\mu$ as
$D \rightarrow D' = D + \Delta$,
$P_\mu \rightarrow P'_\mu = P_\mu + \pi_\mu$,
and
$K_\mu \rightarrow K'_{\mu} = K_\mu + \kappa_\mu$.
In what follows, we consider two cases.
One is the ordinary theory respecting the Poincar\'e symmetry.
The other is analogous to the twister algebra~\cite{penrose},
which does not respect the translational symmetry. 
For later convenience, we introduce the three following sets of conformal generators as
%
%\begin{align*}
$G = \{ D', P'_\mu, K'_\mu, L'_{\mu \nu} \},
%
%\qquad
G_{X} = \{ D, P_\mu, K_\mu, L_{\mu \nu} \}, 
%
%\qquad 
G_{Y} = \{ \Delta, \pi_\mu, \kappa_\mu, s_{\mu \nu} \}$.
%
%\end{align*}

% Ordinary
\subsubsection{Ordinary theory}

In an ordinary context, we assume that a physical state has translational symmetry,
in which case we choose $\pi_\mu$ as vanishing.
At the expense of the vanishing $\pi_\mu$, we cannot represent $\kappa_\mu$ using an operator in the internal degrees of freedom only,
but rather the direct product of the operator and the spacetime variable.
Recalling that $K_\mu$ in (\ref{eq:D=QP}) can be rewritten as $Q^\nu L_{\nu \mu} + Q_\mu (D - \frac{\rm i}{2} d)$,
we find it natural to choose $\kappa_\mu$ as a form analogous to $K_\mu$ where
$L_{\mu \nu}$ and $D$ in $K_\mu$ are replaced by $s_{\mu \nu}$ and $\Delta$,
respectively.
Actually, we have~\cite{lagu,efim}
\begin{align}
\forall g \in G, \; [ \Delta, \, g] = 0, \qquad \pi_\mu \equiv 0, \qquad \kappa_\mu = 2 ( Q_\mu \Delta + Q^\nu s_{\nu \mu}).
\label{eq:Deltag=0}
\end{align}
Note that $\Delta$ plays a similar role to $d$ in $K_\mu$, in the sense that $d$ satisfies
$\forall g \in G, \; [d, \, g] = 0$.
Although we can apply the choice of (\ref{eq:Deltag=0}) to any type of spin structure, such as spinor, vector, and tensor,
a physical meaning (and also a geometrical one) of the invariance of the conformal algebra under
\begin{align}
( D' , P'_\mu, K'_\mu, L'_{\mu \nu} ) \leftrightarrow ( - D', K'_\mu, P'_\mu, L'_{\mu \nu} )
\label{eq:DPKL}
\end{align}
seems to be unclear.

Now, we deal with the conformally invariant integral transform.
We first examine whether or not
the solution to $\mathcal H$ such that $[ \mathcal H, \, x] = 0$ for all $x \in G_X$ satisfies a particular solution such that
$[ \mathcal H, \, z] = 0$ for all $z \in G$.
If it satisfied, the conditions $[ \mathcal H, \, K_\mu] = 0 = [ \mathcal H, \, \kappa_\nu]$ would lead to
$X_{\mu \nu} = 0$ by the Jacobi relation,
where $X_{\mu \nu} = [ \mathcal H, \, [ K_\mu, \, \kappa_\nu]]$.
Applying $[ P_\rho, \cdot]$ to $X_{\mu \nu}$ and using $[ \mathcal H, \, P_\mu] = 0$ and
$[ K_\mu, \, \kappa_\nu] = - (Q \cdot Q) [P_\mu, \, \kappa_\nu] - 2 {\rm i} Q_\mu \kappa_\nu$,
we finally get $0 = [ \mathcal H, \, Q_\mu] s_{\rho \nu} - ( \mu \leftrightarrow \rho)$ and $[ \mathcal H, \, Q_\rho] \Delta = 0$ ($n \neq 2$).
These relations imply that we have $[ \mathcal H, \, Q_\mu] \mathfrak c = 0$,
where $\mathfrak c$ represents the Casimir operator~\cite{lagu}.
In the irreducible representation, $\mathfrak c$ should be proportional to the identity,
so that $[ \mathcal H, \, Q_\mu ] = 0$ is required.
Recalling that we have $[ \mathcal H, \, P_\mu] = 0$, we find that
$\mathcal H$ turns out to be a constant map, a trivial case.
Thus, we have found that in
the process of obtaining a nontrivial (particular) solution to $\mathcal H$ such that $[ \mathcal H, \, z] = 0$ for all $z \in G$,
we cannot impose the condition $[ \mathcal H, x] = 0 = [\mathcal H, y]$ for all $x \in G_X, y \in G_Y$.
% is rather involved,
Due to the rather involved procedure for solving $\mathcal H$,
as is somewhat analogous to the situation in obtaining the conformal spinor~\cite{conf_spinor},
where multiple spinors having different conformal degree $d$ in (\ref{eq:D=QP}) are required,
we do not go on further to explore the general solution to $\mathcal H$.

% Bhabha theory
\subsubsection{Bhabha theory}

In the Bhabha theory~\cite{bhabha} ($n=4$, although the number $n$ can be generalized),
$s_{\mu \nu}$ can be decomposed as $s_{\mu \nu} = {\rm i} [ \beta_{\mu}, \, \beta_{\nu}]$,
in which $[\beta_\mu, s_{\nu \rho} ] = {\rm i} ( g_{\mu \nu} \beta_{\rho} - (\nu \leftrightarrow \rho) )$ to guarantee the commutation relation for $s_{\mu \nu}$,
and the minimum polynomial (not the eigenpolynomial) for ${\rm i} \beta_0, \beta_j \; (j=1,2,3)$ is given by
\begin{align}
f (x) = \prod_{k=0}^{2s} [ x - (s - k) ],
\label{eq:minimum}
\end{align}
where $s \in \{ \frac{1}{2}, 1, \ldots \}$ represents the spin.
The Bhabha theory is characterized by not only $s$ but also another parameter $s'$ such that $s' = s, s-1, s-2, \ldots, \frac{1}{2}$ (or $0$).
Note that for $s = \frac{1}{2}$, $s'$ is allowed to be $\frac{1}{2}$ only.
In the case of $s' = s$, we can choose $\Delta, \pi_\mu$, and $\kappa_\mu$ as~\cite{kuwata}
\begin{align*}
\Delta = {\rm i} \beta_5, \qquad \pi_\mu = \Lambda_+ \beta_\mu, \qquad \kappa_\mu = \Lambda_- \beta_\mu
\qquad \left( \Lambda_\pm = \mathbbm 1 \pm [ \beta_5, \cdot ] \right),
%\qquad s_{\mu \nu} = {\rm i} [ \beta_\mu, \, \beta_\nu],
\end{align*}
where %$\Delta_\pm = \mathbbm 1 \pm [ \beta_5, \cdot ]$, with 
$\beta_5$, satisfying $[ \beta_5, s_{\mu \nu} ] = 0$,
$[ \beta_\mu, \, [ \beta_\nu, \, \beta_5 ]] = g_{\mu \nu} \beta_5$, and
$[ \beta_5, \, [ \beta_5, \, \beta_\mu] ] = \beta_\mu$, is given by using the totally antisymmetric Levi-Civita tensor $\epsilon^{\mu \nu \rho \sigma}$ as
\begin{align*}
\beta_5 = - \frac{\rm i}{N} \epsilon^{\mu \nu \rho \sigma} \beta_\mu \beta_\nu \beta_\rho \beta_\sigma, \qquad N = 2 (s+1).
\end{align*}
Although the value of $N$ is well known for $s = \frac{1}{2}, 1$~\cite{kemmer},
it is hardly known for $s > 1$.
It is rather a complicated task to calculate $N$ for a higher spin,
because as $s$ increase, the order of $f(x)$ in (\ref{eq:minimum}) tends to increase.
Note that as contrasted with the ordinary theory, the invariance of the conformal algebra in $G_Y$ under
$( \Delta, \pi_\mu, \kappa_\mu, s_{\mu \nu} ) \rightarrow ( - \Delta, \kappa_\mu, \pi_\mu, s_{\mu \nu} )$
is realized under $(\beta_\mu, \beta_5) \rightarrow (\beta_\mu, - \beta_5)$ or %$\epsilon_{\mu \nu \rho  \sigma} \rightarrow ( - \epsilon_{\mu \nu \rho \sigma})$.
\begin{align}
\epsilon_{\mu \nu \rho  \sigma} \rightarrow ( - \epsilon_{\mu \nu \rho \sigma}).
\label{eq:e->-e}
\end{align}
Note also that for $s = \frac{1}{2}$, we have
\begin{align*}
\{  2 \Delta, \pi_\mu + \kappa_\mu, \pi_\mu - \kappa_\mu, 2 s_{\mu \nu} \} 
= \{ {\rm i} \gamma_5, \gamma_\mu, \gamma_5 \gamma_\mu, {\rm i} \gamma_\mu \gamma_\nu \; (\mu \neq \nu) \},
\end{align*}
which is in $\mathfrak s \mathfrak u (2,2)$~\cite{penrose},
and is isomorphic to $\mathfrak s \mathfrak o (4,2)$~\cite{haag}.
However, this kind of algebra is rarely called a conformal algebra,
which may be due to the lack of translation symmetry $\pi_\mu = 0$.

To recover the translation invariance in the Bhabha theory,
a reasonable approach is to introduce a physical state $| \psi_{\rm phys} \rangle$ such that
$\pi_\mu | \psi_{\rm phys} \rangle = 0$,
as in the case of the (Lorentz) gauge condition $( \partial \cdot A) | \psi_{\rm phys} \rangle = 0$, rather than $\partial \cdot A = 0$ itself.
Suppose that $| \psi_{\rm phys} \rangle$ can be written using the projector ${\sf P}$ as $| \psi_{\rm phys} \rangle = {\sf P} | \psi \rangle$,
where $| \psi \rangle$ represents an element of the Hilbert space for $G$.
Then we will find that the relation $\pi_\mu | \psi_{\rm phys} \rangle = 0$ holds, provided that ${\sf P}$ satisfies the following condition:
\begin{align}
\beta_5 {\sf P} = s {\sf P}.
\label{eq:bP=sP}
\end{align}
Let $| \psi_{\rm un} \rangle := \pi_\mu {\sf P} | \psi \rangle$.
Using (\ref{eq:bP=sP}) and $[ \beta_5, \, \pi_\mu] = \pi_\mu$,
we obtain $\beta_5 | \psi_{\rm un} \rangle = (s+1) | \psi_{\rm un} \rangle$.
This relation
implies that if $| \psi_{\rm un} \rangle \neq 0$, $(s+1)$ would be one of the eigenvalues of $\beta_5$.
However, the minimum polynomial for $\beta_5$ is given by (\ref{eq:minimum}), as in the case for ${\rm i} \beta_0$ and $\beta_j \; (j=1, 2, 3)$,
so that $\beta_5$ cannot take an eigenvalue of $(s+1)$.
Hence, we should require that $| \psi_{\rm un} \rangle = 0$,
indicating the relation $\pi_\mu {\sf P} = 0$.

So far, we have found that $\pi_\mu | \psi_{\rm phys} \rangle = 0$ is guaranteed by (\ref{eq:bP=sP}).
Such ${\sf P}$ is given by the projection onto the eigenspace of $\beta_5$, with its eigenvalue $s$, as%so that we can write ${\sf P}$ as
\begin{align}
{\sf P} = \frac{1}{f' (s)} \frac{ f (\beta_5) }{\beta_5 - s},
\label{eq:P=1/f}
\end{align}
where $f'(x)$ represents the derivative of $f(x)$.
In this case, 
${\sf P}$ satisfies the two following relations~\cite{kuwata}:
\begin{align}
{\rm dim} \, ( {\sf P} ) = 2s + 1, \qquad \langle s^2 \rangle {\sf P} = s (s+1) {\sf P},
\label{eq:dimP}
\end{align}
where $ \langle s^2 \rangle = \frac{1}{2} \sum_{i,j=1}^3 s_{i j} s^{ij}$.
Thus, we can interpret $| \psi_{\rm phys} \rangle$ as the spin $s$ state (of a massive particle).
Conversely, if we assume that the projection ${\sf P}$ in the Bhabha theory satisfies both relations in (\ref{eq:dimP}),
then we find that $s' = s$ is required (hence the existence of $\beta_5$ is guaranteed),
and that there are two ${\sf P}$'s: one is given by (\ref{eq:P=1/f}) and the other is such that
\begin{align*}
\widetilde{\sf P} = \left . {\sf P} \right |_{s \rightarrow -s} \; \left ( =  \left . {\sf P} \right |_{\beta_5 \rightarrow - \beta_5} \right),
\end{align*}
so that $\widetilde{P}$ can also be given by transforming ${\sf P}$ under (\ref{eq:e->-e}).
Considering that for
$s = \frac{1}{2}$,
we have ${\sf P} = \frac{1}{2} ( 1 + \gamma_5)$ and $\widetilde{\sf P} = \frac{1}{2} ( 1 - \gamma_5)$,
we find that ${\sf P}$ and $\widetilde{\sf P}$ represent the right-handed chirality and the left-handed one for a Dirac particle, respectively.
In this sense,
${\sf P}$ and $\widetilde{\sf P}$ for $s > \frac{1}{2}$ can be regarded as the generalization of the chilarity.

Before proceeding further,
some may point out that the condition (\ref{eq:bP=sP}) is so restrictive that we should generalize it to such that
\begin{align}
\beta_5 {\sf P} = g (s) {\sf P} \qquad \left( g: \mathbb R \rightarrow \mathbb C \right),
\label{eq:bP=gP}
\end{align}
provided that $g(s)$ satisfies
\begin{align}
g (s) + 1 \notin {\rm Spect} \, (\beta_5), % = \{ s, s-1, \ldots, -s \}.
\label{eq:gs+1}
\end{align}
where ${\rm Spect} \, (\beta_5)$ represents the spectrum of $\beta_5$ as $\{ s, s-1, \ldots, -s \}$.
However, we will find that $g (s) = s$ as follows.
The repeated application of $\beta_5$ to (\ref{eq:bP=gP}) yields
$\beta_5^{\, n} {\sf P} = g(s)^n {\sf P}$ ($n \in \mathbb N$),
so that we have
$f (\beta_5) {\sf P} = f (g (s) ) {\sf P}$,
which, together with $f (\beta_5) = 0$, implies $f (g(s) ) = 0$.
Thus, it is required that $g(s) \in {\rm Spect} \, (\beta_5)$,
which, together with (\ref{eq:gs+1}), leads to $g(s) = s$.

Now we obtain the integral transform $\mathcal H$ such that $[ \mathcal H, g] = 0$ for all $g \in G$.
As opposed to the choice (\ref{eq:Deltag=0}),
it is almost apparent that the spacetime conformally invariant integral transform (denoted by $\mathcal H_0$),
that is, $[\mathcal H_0, \, x] = 0$ for all $x \in G_X$, can still remain a particular solution to $\mathcal H$,
owing to $[ x,  \, y] = 0$ for all $x \in G_X$ and $y \in G_Y$.
In this case, we can choose $\mathcal H_0$ as
such that satisfies $[ \mathcal H_0, \, y] = 0$ for all $y \in G_Y$.
Moreover, the general solution to $\mathcal H$ can be given by multiplying $\mathcal H_0$ with $\mathcal C$ (from the left or the right)
such that $[\mathcal C, \, g] = 0$ for all $g \in G$.
(Note that as long as $\mathcal C$ can be given by a function of the generators in $G$,
as in the case of the Casimir operator,
then it follows that $[ \mathcal H_0, \, \mathcal C] = 0$ by $[ \mathcal H_0, \, g] = 0$ for all $g \in G$.)
The convenient choice of $\mathcal C$ is to adopt the Casimir operator of the corresponding conformal group.
Note that
the $\mathcal C$ in the ordinary spin-dependent conformal theory
is given by the polynomial of $\Delta, s_{\mu \nu} s^{\mu \nu}$,
and $s_{\mu \nu} \star s^{\mu \nu}$~\cite{lagu} ($\star s_{\mu \nu}$ represents the Hodge dual of $s_{\mu \nu}$ as
$\star s_{\mu \nu} = \frac{1}{2} \epsilon_{\mu \nu \rho \sigma} s^{\rho \sigma}$),
all of which are spacetime independent,
while
%and $\epsilon_{\mu \nu \rho \sigma} s^{\mu \nu} s^{\rho \sigma}$~\cite{lagu},
the $\mathcal C$ in the Bhabha conformal theory is composed of spacetime-dependent terms,
such as the spin-orbit interaction
$s_{\mu \nu} L^{\mu \nu}$.
For $n=4$, we have three $\mathcal C$'s, denoted by $\mathcal C^{(i)} \; (i=2,3,4)$.
Let $\mathcal C_X^{(i)}$ and $\mathcal C_Y^{(i)}$ be such that are given by replacing $g \in G$ in $\mathcal C^{(i)}$ by the corresponding $x \in G_X$ and $y \in G_Y$,
respectively.
Then, 
we obtain $\mathcal C_Y^{(i)}$ ($C^{(i)}$ is conversely given by replacing $y \in G_Y$ with the corresponding $g \in G$) as~\cite{so:casimir,casimir}
%where $\mathcal C_2 = - D^{\prime 2} + \frac{1}{2} L'_{\mu \nu} L^{\prime \mu \nu} - \frac{1}{2} \{ K'_\mu, \, P^{\prime \mu} \}$,
%
\begin{align*}
\mathcal C_Y^{(2)} &= - \Delta^2 + \frac{1}{2} s_{\mu \nu} s^{\mu \nu} + \frac{1}{2} t_\mu^{\; \; \mu}, \\ 
\mathcal C_Y^{(3)} & = ( \Delta s_{\mu \nu} + t_{\mu \nu} ) \star \! s^{\mu \nu}, \\
\mathcal C_Y^{(4)} & = - \frac{1}{16} ( s_{\mu \nu} \star \! s^{\mu \nu} )^2 + \frac{1}{2} w_{\mu \nu} w^{\mu \nu} - \frac{1}{2} \{ \mathfrak k_\mu, \, \mathfrak p^\mu \},
\end{align*}
where $t_{\mu \nu}, w_{\mu \nu}$, $\mathfrak k_\mu$, and $\mathfrak p_\mu$ are given by
\begin{align*}
t_{\mu \nu} = \{ \kappa_\mu, \, \pi_\nu \}, \qquad
w_{\mu \nu} = \Delta \star \! s_{\mu \nu} + \frac{1}{2} \star \! t_{\mu \nu}, \qquad
( \mathfrak k_\mu, \mathfrak p_\mu) =  \star s_{\mu \nu} ( \kappa^\nu, \pi^\nu).
\end{align*}
Note that under (\ref{eq:e->-e}),
we have $( t_{\mu \nu}, w_{\mu \nu}, \mathfrak k_\mu, \mathfrak p_\mu) \rightarrow (t_{\nu \mu}, w_{\mu \nu}, - \mathfrak p_\mu, - \mathfrak k_\mu)$,
so that we find that all the $\mathcal C_Y^{( i )}$'s are invariant under (\ref{eq:e->-e}).

It may be interesting to examine how $\mathcal C$ plays a role in the "physical" state ${\sf P} | \psi \rangle$.
Neglecting the terms proportional to the identity in $\mathcal C$
[they come from $\mathcal C_X$ and $\mathcal C_Y$, where we have
$\mathcal C_X^{(2)} = \frac{d}{2} ( \frac{d}{2} - n)$,
$\mathcal C_X^{(3)} = 0 = \mathcal C_X^{(4)}$; and
$\mathcal C_Y^{(2)} = 3 s (s+2)$,
$\mathcal C_Y^{(3)} = 4s (s+1) (s+2)$,
$\mathcal C_Y^{(4)} = 3 s (s+1)^2 (s+2)$],
we can write $\mathcal C^{(i)} {\sf P}$ as
\begin{align*}
\mathcal C^{(2)}  {\sf P} = 2 \left ( -  {\rm i} s D + I + \beta_\mu P^\mu \right)  {\sf P}, \qquad
\mathcal C^{(3)} {\sf P} = 4 (s+1) \left ( -  {\rm i} s  D +  {\rm i}  J +  \beta_\mu W^\mu \right ) {\sf P},
\end{align*}
where $I = \frac{1}{2} s_{\mu \nu} L^{\mu \nu}$,
%$\not{\! P} = 2 \beta_\mu P^\mu$,
%
$J = \frac{1}{2} \star s_{\mu \nu} L^{\mu \nu}$,
and
%$\not{\! W} = 2 \beta_\mu W^\mu$, with
$W_\mu = \frac{1}{s+1} \star s_{\mu \nu} P^\nu$ (Pauli-Lubanski pseudo vector).
Considering that
for $s' = s$, we have~\cite{kuwata}
\begin{align}
\star s_{\mu \nu} {\sf P} = - {\rm i} s_{\mu \nu} {\sf P},
\label{eq:sP=-isP}
\end{align}
we obtain $I {\sf P} = {\rm i} J {\sf P} $.
We also get $\beta_\mu P^\mu {\sf P} =  \beta_\mu W^\mu {\sf P}$
by the four following relations:
 (\ref{eq:sP=-isP}),
$[ \beta_5, \, [ \beta_5, \, \beta_\mu]] = \beta_\mu$,
$\pi_{\mu} {\sf P} = 0$,
and $\beta_\mu \beta^\mu = s (s+2) - \beta_5^{\, 2}$ (for $s' = s$)~\cite{kuwata}.
%
%together with
%$[ \beta_5, \, [ \beta_5, \, \beta_\mu]] = \beta_\mu$
%and
%$\pi_{\mu} {\sf P} = 0$,
%$[ \beta_5, \, [ \beta_5, \, \beta_\mu]] = \beta_\mu$,
%together with
%the relation $\beta_\mu \beta^\mu = s (s+2) - \beta_5^{\, 2}$ (for $s' = s$)~\cite{kuwata}.
%and recalling that $[ \beta_5, \, [ \beta_5, \, \beta_\mu]] = \beta_\mu$ and $\pi_{\mu} {\sf P} = 0$,
%we get $\not{\! W} {\sf P} = (s+1) \not{\! P} {\sf P}$.
%
Thus, $\mathcal C^{(3)} {\sf P}$ is simply related to $\mathcal C^{(2)} {\sf P}$ through the relation
\begin{align*}
\mathcal C^{(3)} {\sf P} = N  \mathcal C^{(2)}  {\sf P},
\end{align*}
which is somewhat analogous to the relation $\mathcal C_Y^{(3)} = \frac{2}{3} N \mathcal C_Y^{(2)}$.
In a similar way,
we can calculate $\mathcal C^{(4)} {\sf P}$,
which is expected to be written using certain $\lambda, \mu \in \mathbb R$ as $\lambda \mathcal C^{(2)} {\sf P} + \mu  ( \mathcal C^{(2)} )^2 {\sf P}$
by the relation $\mathcal C_Y^{(4)} = \mathcal C_Y^{(2)} + \frac{1}{3} ( \mathcal C_Y^{(2)} )^2$.
However, the actual calculation reveals that some other terms are necessary.
For $s=\frac{1}{2}$, for example,
in which $\lambda = 1$ and $\mu = \frac{1}{4}$ ($\lambda$ and $\mu$ may depend on the value of $s$),
we should add two terms:
one is proportional to $\beta_\mu P^\mu {\sf P}$, and the other is proportiona to $C_X^{(2)} {\sf P}$.
%

% Discussion
\section{Discussion}

This section discusses the spin-orbit duality in connection with the spin-dependent conformal transformation.
In the Bhabha theory, it is almost apparent that the conformal symmetry is invariant under the exchange of $s_{\mu \nu} \leftrightarrow L_{\mu \nu}$.
However, if we respect the Poincar\'e symmetry, as in the ordinary spin-dependent conformal transformation,
there seems no apparent spin-orbit duality $s_{\mu \nu} \leftrightarrow L_{\mu \nu}$.
%In the Bhabha theory, it is almost apparent that we have the spin-orbit duality $s_{\mu \mu} \leftrightarrow L_{\mu \nu}$.
%If the theory under consideration is invariant under the Hodge product, then
Instead, we may have another type of spin-orbit duality such that
$(s_{\mu \nu}, L_{\mu \nu} ) \rightarrow (\star L_{\mu \nu}, \star s_{\mu \nu} )$,
where $\star X_{\mu \nu} = \frac{1}{2} \epsilon_{\mu \nu \rho \sigma} X^{\rho \sigma}$.
Such a spin-orbit duality,
together with $P_\mu \rightarrow P_\mu$,
was discussed extensively in Ref.~\cite{filippas},
where the spin supplementary condition $s_{\mu \nu} P^\nu = 0$ was employed to eliminate three undetermined degrees of freedom for $s_{\mu \nu}$
from the energy-momentum conservation
$T^{\mu \nu}_{ \; \;  \; ; \nu} = 0$.
Although we employ the spin supplementary condition for relativistic spinning (vector) particles~\cite{corinadesi},
it may be too restrictive for the Bhabha particle in the following sense.
To begin with, by (\ref{eq:sP=-isP}),
together with $s_{\mu \nu} P^\nu = 0$, we have
\begin{align}
W_\mu {\sf P} = 0.
\label{eq:WP=0}
\end{align}
Furthermore, under the substitution $\epsilon_{\mu \nu \rho \sigma} \rightarrow - \epsilon_{\mu \nu \rho \sigma}$,
we have
\begin{align}
- W_\mu \widetilde{\sf P} = 0.
\label{eq:-WP=0}
\end{align}
For $s = \frac{1}{2}$, in which ${\sf P} + \widetilde{\sf P} = 1$,
we obtain $W_\mu = 0$ by (\ref{eq:WP=0}) and (\ref{eq:-WP=0}).
Considering that $W \cdot W = \frac{3}{4} P \cdot P$ ($s = \frac{1}{2}$),
we find that $P \cdot P = 0$ by $W_\mu = 0$.
Thus, the condition $s_{\mu \nu} P^\nu = 0$ indicates that the mass for a free Dirac particle ($s=\frac{1}{2}$) is vanishing.

It may be intriguing to examine whether or not the transformation of (\ref{eq:sP=-isP}) under $s_{\mu \nu} \rightarrow L_{\mu \nu}$ still holds.
This statement is not trivial because (\ref{eq:sP=-isP}) holds for the case of $s' = s$.
To begin with, we show that $\beta_5 \rightarrow 0$ ($s_{\mu \nu} \rightarrow L_{\mu \nu}$).
Recalling that $\beta_5$ can be rewritten as $\frac{\rm i}{4 N} \epsilon^{\mu \nu \rho \sigma} s_{\mu \nu} s_{\rho \sigma}$,
we get $\beta_5 \rightarrow \frac{\rm i}{4 N} \epsilon^{\mu \nu \rho \sigma} L_{\mu \nu} L_{\rho \sigma}$,
which is identically vanishing by $L_{\mu \nu} \star L^{\mu \nu} \equiv 0$.
As a consequence, ${\sf P}$ is transformed as
\begin{align*}
{\sf P} \rightarrow \hat{\sf P}  = \lim_{x \rightarrow 0} \frac{1}{ ( 2 \ell )! } \frac{1} {x - \ell} \prod_{k=-\ell}^{\ell} ( x - k) 
= \begin{cases} 1 & ( \ell = 0), \\
0 & ( \ell = 1,2,\ldots),
\end{cases}
 %  \qquad  \left( \ell = 0, 1, 2, \ldots \right),
\end{align*}
where we have replaced $s$ in (\ref{eq:minimum}) by $\ell$,
which represents the maximum eigenvalue of $L_{\mu \nu}$,
that is, $L_{\mu \nu}$ takes an eigenvalue in $\{\ell, \ell-1, \ldots, - \ell \}$.
As contrasted with the eigenvalue of $s_{\mu \nu}$, the eigenvalue of half an integer is not allowed for $L_{\mu \nu}$.
%Thus we have $\hat{\sf P} = 1$ ($\ell = 0$) and $0$ ($\ell = 1,2,\ldots$).
%
Eventually, we obtain the following relation:
\begin{align}
\star L_{\mu \nu} \hat{\sf P} = - {\rm i} L_{\mu \nu} \hat{\sf P},
\label{eq:LP=-iLP}
\end{align}
which holds for all $\ell \in \{ 0,1,2,\ldots \}$,
although it is a trivial relation ($0 = 0$).

It is more instructive to transform the spin relation with no projection ${\sf P}$ (because $\hat{\sf P}$ is almost trivial).
We give one such example~\cite{kuwata}
\begin{align}
\frac{1}{2} s_{\mu \nu} s^{\mu \nu} = 2 s (s+1) - ( s^2 - \beta_5^{\, 2} ) \qquad (s' = s),
\label{eq:ss=2s}
\end{align}
which is transformed to $\frac{1}{2} L_{\mu \nu} L^{\mu \nu} = 2 \ell (\ell+1) - \ell^2$, not $2 \ell (\ell+1)$.
Using the ordinary relation $\frac{1}{2} L_{ij} L^{ij} = \ell (\ell + 1)$,
we expect that the corresponding orbital angular momentum relation to (\ref{eq:ss=2s}) dictates that
%While (\ref{eq:ss=2s}), together with (\ref{eq:sP=-isP}) and $\beta_5 {\sf P} = s {\sf P}$, leads to $\frac{1}{2} s_{ij} s^{ij} {\sf P} = s (s+1) {\sf P} = s_{0i} s^{0i} {\sf P}$
%(for $s= \frac{1}{2}$, ${\sf P}$ can be eliminated, and its validity can be checked using the gamma matrices),
%
%the orbital angular momentum relation dictates that [using $\frac{1}{2} L_{ij} L^{ij} = \ell (\ell + 1)$]
%
\begin{align}
L_{0i} L^{0i} = \ell.
\label{eq:LL=ell}
\end{align}
%
%which is not $\ell (\ell + 1)$, as expected from the spin relation. %, where use has been made of $\frac{1}{2} L_{ij} L^{ij} = \ell (\ell + 1)$.
%
Compared with the spin relation, where
$s_{0i} s^{0i} {\sf P} =  s (s+1) {\sf P}$ by (\ref{eq:sP=-isP}), (\ref{eq:ss=2s}), and $\beta_5 {\sf P} = s {\sf P}$
(for $s= \frac{1}{2}$, ${\sf P}$ can be eliminated),
the relation (\ref{eq:LL=ell}) seems, at first, intriguing.
However, we will discuss the validity of (\ref{eq:LL=ell}) elsewhere.

%% Conclusion
\section{Conclusion}

We have found that, based on the Bhabha theory, we can write the spin-dependent conformal invariant integral transform as the product of
the spin-independent integral transform (whose integral kernel is the dilatation invariant Riesz potential) and the Casimir operators of the corresponding conformal group.
While the ordinary theory respects the Poincar\'e symmetry, we introduce the intrinsic momentum operator $\pi_\mu$,
as in the twister algebra,
so that the Casimir operator turns out to be spacetime dependent.
% (such as spin-orbit coupling and Pauli-Lubanski pseudo vector).
%
To accommodate the translational invariance of a physical state $| \psi_{\rm phys} \rangle$, it should be annihilated by the $\pi_\mu$,
to find that the $| \psi_{\rm phys} \rangle$ represents the spin-$s$ state in the sense of (\ref{eq:dimP}).
Compared with the ordinary theory,
the Bhabha theory can deal with the spin-orbit duality more easily,
due to the conformal algebra invariance under $G_X \ni x \leftrightarrow y \in G_Y$.

%% Ack
\section*{Acknowledgments}

S. K.  is indebted to Y. Nasuda for valuable discussions.
This work was partially supported in part by SATAKE TECHNICAL FOUNDATION and is supported in part by JSPS KAKENHI Grant No. JP23K03251.

%% Ref

%% End of file
\end{document}